\documentclass[a4paper,11pt]{article}

\usepackage{jinstpub}
\usepackage{lineno,hyperref}
\usepackage{hhline}
\usepackage{array}
\usepackage{multirow}
\usepackage{booktabs}
\usepackage{tabularx,hhline}
\usepackage{float}
\usepackage{amsmath}
\usepackage{gensymb}
\usepackage{color,soul} 
\usepackage{mathtools, cuted}
\usepackage{lipsum, color}
\usepackage{lipsum}
\usepackage{graphicx}
\usepackage{subcaption}
\usepackage{textcomp}

\hypersetup{
    colorlinks=true,       
    linkcolor=blue,          
    citecolor=green,        
    filecolor=magenta,      
    urlcolor=cyan           
}
\graphicspath{{images/}{./}}
\modulolinenumbers[5]

\title{SiPM behaviour under continuous light}

\author[1,a]{A. Nagai,\note{Corresponding author.}}
\author[a]{C. Alispach,}

\author[a]{D. della Volpe,}
\author[a]{M. Heller,}

\author[a]{T. Montaruli,}
\author[a]{S. Njoh,}

\author[a]{I. Troyano-Pujadas,}

\author[a]{Y. Renier}

\affiliation[a]{D\'epartment de physique nucl\'eaire et corpusculaire, Universit\'e de Gen\`eve, 24 Quai E. Ansermet, CH-1211, Switzerland}

\emailAdd{Andrii.Nagai@unige.ch}

\abstract{This paper reports on the behaviour of Silicon Photomultiplier (SiPM) detectors under continuous light.
Usually, the bias circuit of a SiPM has a resistor connected in series to it, which protects the sensor from drawing too high current.
This resistor introduces a voltage drop when a SiPM draws a steady current, when illuminated by constant light. This reduces the actual SiPM bias and then its sensitivity to light. As a matter of fact, this effect changes all relevant SiPM features, both electrical (i.e. breakdown voltage, gain, pulse amplitude, dark count rate and optical crosstalk) and optical (i.e. photon detection efficiency).
To correctly operate such devices, it is then fundamental to calibrate them under various illumination levels.

In this work, we focus on the large area ($\sim$1~cm$^{2}$) hexagonal SiPM S10943-2832(X) produced by Hamamatsu HPK for the camera of a gamma-ray telescope with 4 m-diameter mirror, called the SST-1M.
We characterize this device under light rates raging from 3~MHz up to 5~GHz of photons per sensor 
at room temperature (T = 25 $^{\circ}$C). 
From these studies, a model is developed in order to derive the parameters needed to correct for the voltage drop effect. This model can be applied for instance in the analysis of the data acquired by the camera to correct for the effect. The experimental results are also compared with a toy Monte Carlo simulation and
finally, a solution is proposed to compensate for the voltage drop.}


\begin{document}
\maketitle
\flushbottom

\section{Introduction}

SiPM detectors have become the preferred photosensors for many applications in high-energy particle and astroparticle physics, and they are very attractive also for LIDAR and medical imaging applications for diagnostic and therapeutic purposes. More traditionally, such applications employ photomultipliers tubes (PMTs) or multi-anode PMTs (MA-PMTs). 
Between the important advantages of SiPMs  there are compactness, speed of response, insensitivity to magnetic fields, high gain, and low operating voltage. Additionally, with respect to PMTs, they offer the possibility to operate under high and continuous illumination without ageing. 
While SiPM devices are commonly used as single photon detectors operating in dark conditions, in this study we  describe how SiPM devices can be used as multi-photon detectors under continuous light (CL). 

As an example application of SiPM in the presence of CL, SiPMs are now being adopted for cameras of Imaging Atmospheric Cherenkov Telescopes (IACTs) dedicated to gamma-ray astronomy. The pathfinder of this technology applied to gamma-ray astronomy has been the FACT telescope \cite{FACT}. Further developments have been performed for the implementation of the Cherenkov Telescope Array (CTA) \cite{CTAWeb,Acharya2017}, in the frame of which  the SST-1M was originally developed \cite{Montaruli:2015xya} with its SiPM-based camera \cite{CameraPaperHeller2017}.  Thanks to their robustness, the CL at which SiPMs can operate is higher than for photomultipliers \cite{Neise:2017ldg}. For IACTs, CL is due to night sky background (NSB), meaning stray light from reflections from ground and light from human induced sources, and light due the presence of high moon. 

Continuous light on a SIPM sensor leads to a steady current flowing through it. 
To prevent such a high current to eventually damage it, a bias resistor, $R_{bias}$, is
connected in series with the SiPM. This current flowing in $R_{bias}$ translates into a voltage drop $V_{drop}$ at the SiPM bias stage thus reducing the current. It is a typical negative feedback loop.
At the same time, the $V_{drop}$ affects the over-voltage, $\Delta V = V_{BD} - V_{PS}$ (where $V_{BD}$ is the breakdown voltage and $V_{PS}$ the bias voltage supplied by the power supplier PS), which impacts most of the SiPM parameters: gain ($G$), timing, photon detection efficiency ($PDE$), dark count rate ($DCR$), optical crosstalk probability ($P_{ct}$) and afterpulse probability ($P_{ap}$).
Therefore, when SiPM devices are used in the presence of CL, all these parameters have to be characterized as a function of the impinging light intensity to correctly define the performance of a sensor. The consequence of the variation of SiPM operation parameters is that the same incoming physics signal produces a different output when observed in different CL levels. 
At the analysis level, the corrections for this effect must be applied to interpret experimental data properly. 
These corrections are important for the operation of SiPM-based gamma-ray cameras and also for applications where sensors are subject to high radiation levels, which induce an increasing dark count rate with increasing integrated dose. 
In this case, the dark count rate can reach such high levels to mimic CL with rates of the order of several MHz ~\citep{GARUTTI}.

In Sec.~\ref{sec:cont_light} we illustrate further the behaviour of SiPM devices under CL. In Sec.~\ref{sec:model} we describe the voltage drop process mathematically and in Sec.~\ref{sec:toy} we describe the implementation of a toy Monte Carlo (MC) for DC coupled electronics systems. After, in Sec.~\ref{sec:valid} we show the comparison of the proposed toy MC model with the simplified analytical calculation and with measurements obtained in the laboratory with a calibrated light source (Sec~\ref{Sec:CalibratedSource}) and with the SST-1M camera and its Camera Test Setup (see Sec.~\ref{Sec:ValidCTS}).

\section{SiPM behaviour under continuous light}
\label{sec:cont_light}

In this section, we describe how SiPM devices can also be used as multi-photon detectors under CL.
In this paper we focus on a DC coupled electronics associated to the SiPM. 

\begin{figure}[hbt]
\centering
\includegraphics[width=0.95\textwidth]{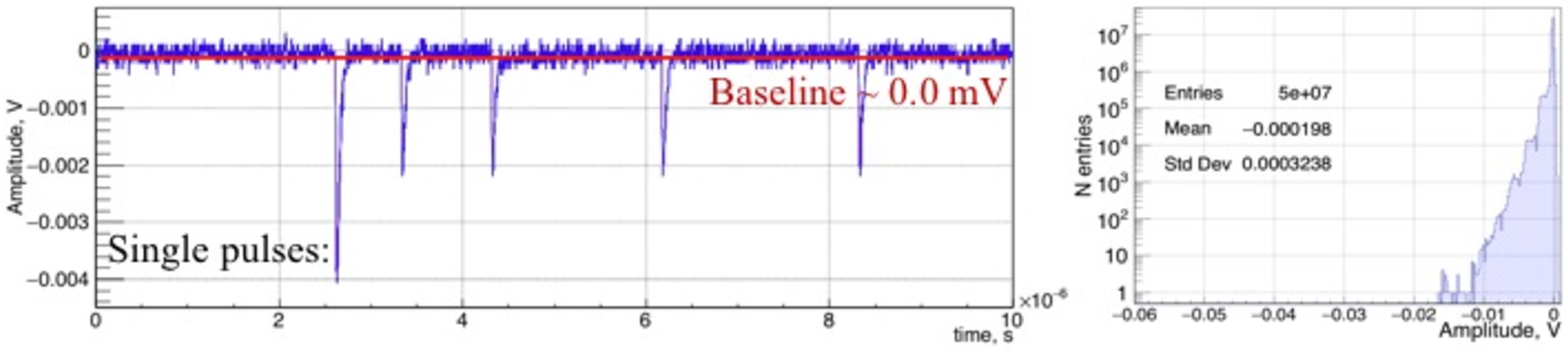}\hfill
\includegraphics[width=0.95\textwidth]{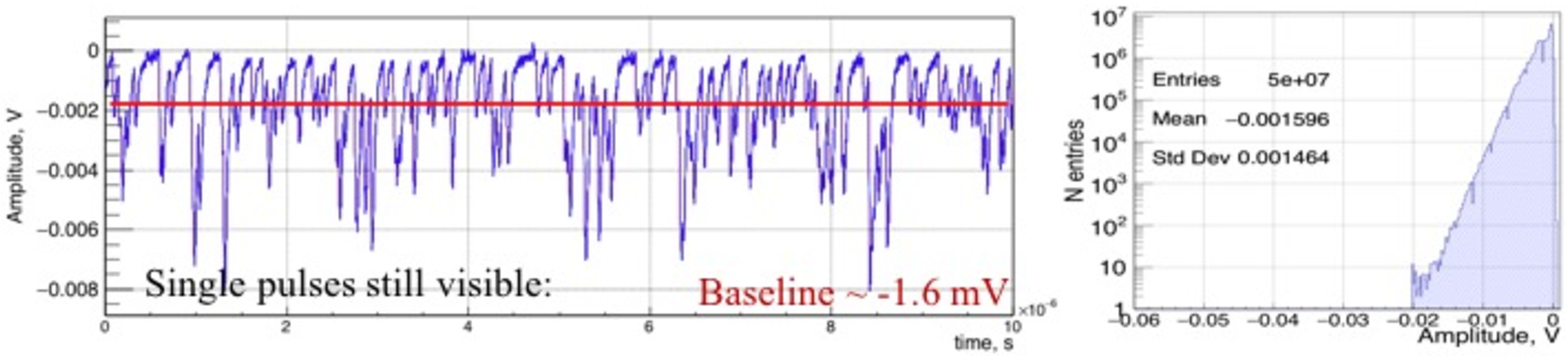} \hfill
\includegraphics[width=0.95\textwidth]{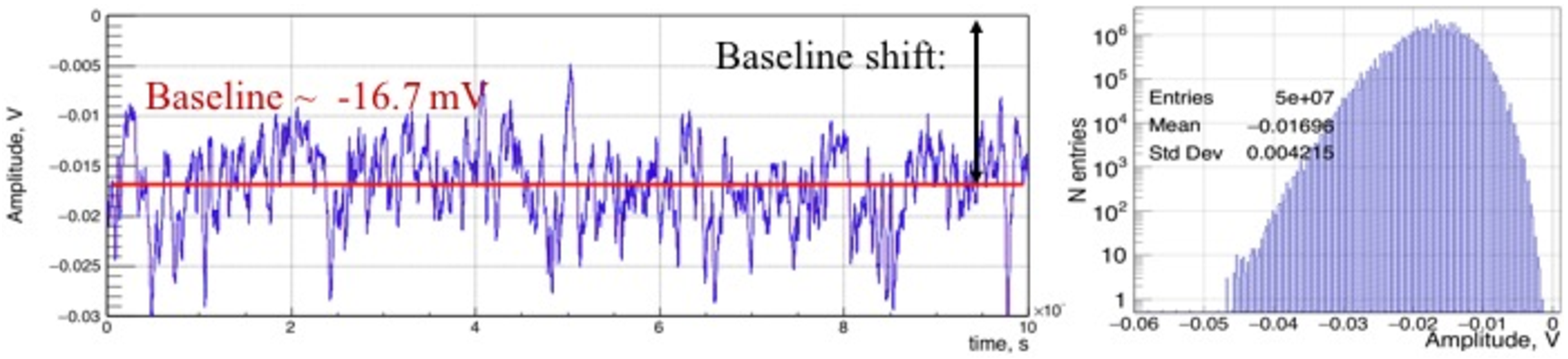}
\caption{From top to bottom: Typical response (waveform) of SiPM in dark conditions (3~MHz $DCR$), under illumination with photon rates of 40.9~MHz and 660~MHz. The amplitude distribution for 10'000 waveforms is shown on the right side for each light intensity level.}
\label{Fig:SiPMWaveforms}
\end{figure}

Typical digitized experimental waveforms obtained in dark conditions and under two levels of CL are presented in Fig.~\ref{Fig:SiPMWaveforms}-left, while the corresponding amplitude $W_{i}$ distribution is shown in Fig.~\ref{Fig:SiPMWaveforms}-right. In dark conditions, the number of generated avalanches  $N_{av}$ can be calculated by simple counting of SiPM pulses. However, above a given level of CL intensity, SiPM pulses become indistinguishable. This happens when the pulse duration of a single photon becomes compatible with the time lapse between 2 photons, i.e. the photon rate. Therefore the counting of SiPM pulses becomes impossible. Nevertheless, under high CL, $N_{av}$ can be approximated by:
\begin{equation}
 \label{Eq:PhotonRate}
N_{av} = \frac{BLS \times \Delta t }{Q_{1pe}}
\end{equation}
where $Q_{1pe}$ is the integral of the single p.e. pulse over its pulse length, $\Delta t$ is the waveform length and $BLS$ is the baseline shift calculated as:
\begin{equation}
    BLS = \frac{1}{S} \left( \sum_{i=0}^{S - 1} W_{i}(\Delta V > 0) - \sum_{i=0}^{S - 1} W_{i}(\Delta V < 0) \right)
\end{equation}
where $W_{i}(\Delta V > 0)$ and $W_{i}(\Delta V < 0)$ are the amplitude values of the experimental waveforms for a given sample $i$ acquired with the SiPM  biased above the breakdown voltage $V_{BD}$ and below, respectively, and $S$ is the number of waveform samples.
It is worth to mention that $W_{i}(\Delta V < 0)$ is dominated by the electronics baseline, while $W_{i}(\Delta V > 0)$ contains also both detected light pulses (if the SiPM is exposed to light) and SiPM correlated noise. 
Correlated noise is due to  afterpulses and cross-talk. There is a given probability that an afterpulse might generate itself other  afterpulses, $P_{ap}$. Therefore,  following the Ref.~\citep{Para}, the total probability that an initial avalanche will be enhanced by afterpulses is:
\begin{equation}
    P_{ap, tot} = P_{ap} + P_{ap}^{2} + P_{ap}^{3} +...=  \frac{P_{ap}}{1-P_{ap}}
\end{equation}
Similarly, for optical cross-talk, with a given probability $P_{ct}$, this enhancement of the cascading effect leads to:
\begin{equation}
    P_{ct, tot} = P_{ct} + P_{ct}^{2} + P_{ct}^{3} +...= \frac{P_{ct}}{1-P_{ct}}
\end{equation}
The $N_{av}$ is the number of avalanches due to the detected photons and augmented by SiPM correlated and uncorrelated noise:
\begin{equation}
    N_{av} = \left( N_{pe} + DCR \times \Delta t \right) \cdot (1 + P_{ct, tot}) \cdot (1 + P_{ap, tot})
\end{equation}
where $N_{pe}$ is the number of detected photons. Therefore, $N_{pe}$, can be calculated as:
\begin{equation}
    N_{pe} = \frac{N_{av} - DCR \times \Delta t}{ (1 + P_{ct, tot}) \times (1 + P_{ap, tot}) },
\end{equation}
and the number of photons $N_{ph}$ can be calculated as:
\begin{equation}
    N_{ph} = \frac{N_{pe}}{PDE} .
\end{equation}

Usually SiPMs are operated in dark conditions and therefore also $PDE$ and noise, namely, $DCR$, $P_{ct}$ and $P_{ap}$, are evaluated in dark conditions.
Nonetheless, SiPMs are usually biased through a RC filter (see e.g. \cite{Hamamatsu}) in order to:
\begin{itemize}
\setlength{\itemsep}{1pt}
\setlength{\parskip}{0pt}
\setlength{\parsep}{0pt}
    \item filter high frequency electronic noise coming from the bias source;
    \item limit the current in order to protect the sensor in case of intense illumination.
\end{itemize}
Due to the presence of the bias resistor $R_{bias}$ and CL, the SiPM parameters deviate from their ``dark" values, meaning their values measured in dark conditions.
As a matter of fact, the voltage drop, $V_{drop}$, induced by the bias resistor  $R_{bias}$ reduces the over-voltage $\Delta V$ as follows:
 \begin{equation}
 \Delta V  = V_{PS} - V_{BD} - V_{drop} =  V_{PS} - V_{BD} - R_{bias} \times I_{SiPM}
 \label{Eq:OverVoltageWithDrop}
 \end{equation}
 where $I_{SiPM}$ is the current generated by the SiPM. 
 \begin{figure}[hbt]
    \begin{center}
        \includegraphics[width=8.5cm, keepaspectratio]{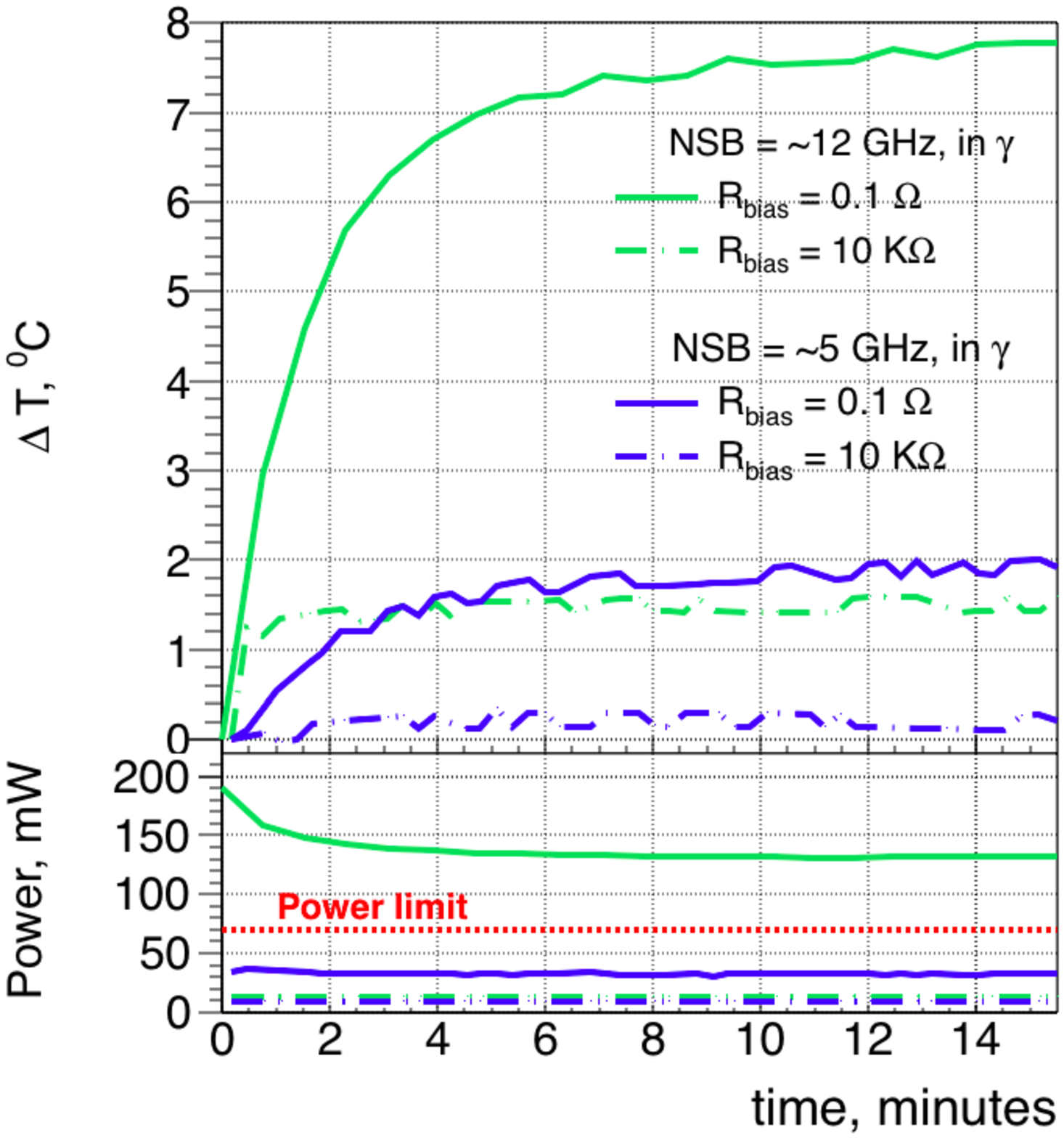}
    \end{center}
    \caption{SiPM temperature (upper panel) and power consumption (bottom panel) at initial $\Delta V$ = 2.8 V and under 12 $\times \ 10^{9}$~ photons/s vs time for $R_{bias}$ = 0.1 $\Omega$ (solid line) and $R_{bias}$ = 10 $k\Omega$ (dashed line). The highest acceptable power consumption for the sensor (provided by the producer) is represented by the red dashed line.}
    \label{Fig:SiPMTemperaturePower}
\end{figure} 
From Eq.~\ref{Eq:OverVoltageWithDrop}, one can conclude that the smaller $R_{bias}$ is, the more stable will be the sensor response. However, having a small $R_{bias}$ also means that high currents can flow through a SiPM, leading to self-heating of the sensor for high CL, as shown in Fig.~\ref{Fig:SiPMTemperaturePower}. If there is a possibility to measure the $I_{SiPM}$, the $V_{drop}$ can be compensated by the feedback system described in Ref.~\cite{FACTNSB}. For other cases, the $I_{SiPM}$ can be calculated analytically or from the toy MC, as shown in next sections.
 

\section{Analytic description of the voltage drop process
\label{sec:model}}

\begin{figure*}[hbt]
    \begin{center}
        \includegraphics[width=1\textwidth,keepaspectratio]{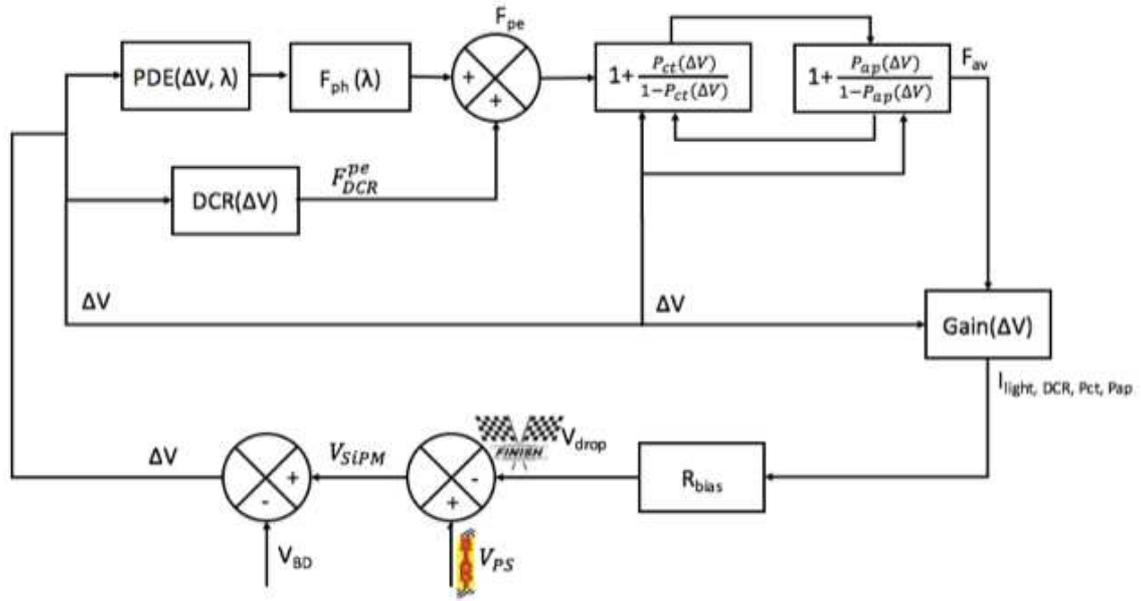}
    \end{center}
    \caption{Representation of the impact of CL on the voltage applied to the sensor. The first loop starts at the START on the bottom, where the $V_{PS}$ is provided to the sensor. The overvoltage is then estimated $\Delta V = V_{PS}- V_{BD}$ and the total rate of avalanches $F_{av}$ is calculated from the injected CL ($F_{ph}$) rate (transformed into p.e. using the $PDE$ and adding uncorrelated noise $DCR$). This p.e. rate is enhanced by correlated noise (i.e. optical crosstalk $P_{ct}$ and afterpulses $P_{ap}$). $F_{av}$ produces the current $I_{light}$ after multiplying by the gain. This current is flowing through the bias resistor $R_{bias}$ and decreases the voltage applied by the user $V_{PS}$ down to $V_{SiPM}$, which is seen by the SiPM. This causes a drop of the values of $G$, $PDE$, $DCR$, $P_{ct}$ $P_{ap}$ and the loop can restart.}
    \label{Fig:ModelIdea}
\end{figure*}

The voltage drop process can be described by the scheme in Fig.~\ref{Fig:ModelIdea}. We adopt in the figure the already defined
voltage supplied by the power supply, $V_{PS}$, the breakdown voltage $V_{BD}$ and the one at the SiPM terminals $V_{SiPM}$. 
The rates of CL expressed in photons and in photo-electrons (p.e.) per unit of time are, respectively, $F_{ph}$ and $F_{pe}$ (where $F_{pe}$ is obtained from $F_{ph}$ at a certain wavelength using the $PDE$). 
The important input parameters for this model are:
\begin{itemize}
\setlength{\itemsep}{1pt}
\setlength{\parskip}{0pt}
\setlength{\parsep}{0pt}
\item the microcell $C_{\mu cell}$ and parasitic $C_{q}$ capacitance's, which determine the SiPM gain, $G= (C_{\mu cell} + C_{q}) \cdot \Delta V/ e$, and therefore allow to convert p.e. to current;
\item the $PDE(\lambda, \Delta V)$ determines the probability that a photon of a given wavelength is detected at a given over-voltage $\Delta V$;
\item $DCR$ determines the rate of thermally generated avalanches at a given $\Delta V$ (i.e. the uncorrelated SiPM noise);
\item The optical crosstalk probability, $P_{ct}$ and the after
pulse probability, $P_{ap}$ (i.e. the correlated SiPM noise). Both probabilities enhance the total rate of avalanches produced by SiPM at a given $\Delta V$;
\item a precise template of the typical normalized SiPM pulse due to 1 p.e. at a given temperature. The SiPM pulse template can be calculated by averaging a given number of normalized single~\footnote{A single p.e. pulse is separated by neighboring pulses by a time interval longer than the sum of the typical SiPM rise and recovery times.} pulses. 
Special care is taken in retrieving the slow component of the pulse. Even if its contribution to the total charge is small (e,g. $<$ 1\%), with rates up to few GHz, neglecting it would lead to large discrepancies on the waveform baseline shift, which is used to derive the level of CL\footnote{CL is provided by NSB in the case of gamma-ray cameras.}. 
See Sec.~\ref{Sec:CalibratedSource} for more information about how to compensate for the slow component of the pulse.
\end{itemize}

Apart from $C_{\mu cell}$, all aforementioned parameters depend on $\Delta V$ and are therefore effected by the voltage drop $V_{drop}$.

The calculation and simulation of the voltage drop requires to calculate dynamically the current generated by the SiPM:
\begin{equation}
        I_{SiPM} = F_{av} \times \frac{(C_{\mu cell} + C_{q})
        \cdot \Delta V }{e}
\label{Eq:SiPMCurrent}
\end{equation}
with $\Delta V$ given by Eq.~\ref{Eq:OverVoltageWithDrop}.The total rate of avalanches produced by the detected light enhanced by correlated and uncorrelated noise $F_{av}$, is given by:
\begin{align}
F_{av} & = \left( DCR + F_{ph} \cdot PDE \right)
ç    \cdot \left( 1 + \frac{P_{ct}}
     {1 - P_{ct}}\right) 
     \cdot \left( 1 + \frac{P_{ap}}{1 - P_{ap}} \right)
     \label{Eq:Favalanches}
\end{align}
Following Ref.~\citep{Nagai:2018ovm}, $PDE$, $P_{ct}$ , $P_{ap}$ and $DCR$ can be parameterized as:
     \begin{align}
     \label{Eq:PDE}
        & PDE = PDE_{max} \cdot P_{G}\\
     \label{Eq:Pxt}
        & P_{ct} = G \cdot P_{h \nu} \cdot P_{G}^{XT} \\
     \label{Eq:Pap}
        & P_{ap} = G \cdot P_{trap} \cdot P_{G}^{P_{AP}} \\
     \label{Eq:DCR}
        & DCR = N_{car} \cdot P_{G}^{P_{DCR}} \cdot e^{b \cdot V_{SiPM}},
    \end{align}
where $PDE_{max}$ is a free parameter, which depends on the SiPM type, the light wavelength and, to some extent, on the temperature; $P_{G}$, $P_{G}^{XT}$, $P_{G}^{P_{AP}}$ and $P_{G}^{P_{DCR}}$ are the average Geiger probabilities for external light of a given wavelength, optical crosstalk, after-pulses and dark pulses, respectively, and $P_{trap}$ is the probability that a carrier is trapped and released after; $P_{h \nu}$ is the probability that a photon is emitted and reaches the high field region of another $\mu$cell and creates
an electron-hole pair; $N_{car}$ is the rate of thermally generated carriers; $b$ is another free parameter describing the increase of $DCR$ with $V_{SiPM}$ due to electrical field effects. 

Eq.~\ref{Eq:SiPMCurrent} becomes non-linear, once taking into account Eq.~\ref{Eq:OverVoltageWithDrop} and Eqs.~\ref{Eq:Favalanches}--\ref{Eq:DCR}. 
In order to simplify Eq.~\ref{Eq:SiPMCurrent}, the Taylor series expansion can be applied to Eg.~\ref{Eq:Favalanches}-\ref{Eq:DCR}. 
However, to achieve a reasonable agreement between the formula and the measured parameters (i.e. $PDE$, $P_{ct}$ , $P_{ap}$ and $DCR$), the Taylor expansion needs a second or even third terms, which eventually leads to a fifth order in Eq.~\ref{Eq:SiPMCurrent}. 

An analytical calculation of Eq.~\ref{Eq:SiPMCurrent} can be done only for the simplified case that CL affects only the SiPM gain $G$ \citep{SST1Melectronics}, while all other parameters (i.e. $PDE$, $DCR$, $P_{ct}$ and $P_{ap}$) are not affected. This allows to express the voltage drop as a function of the CL rate as follows:
\begin{equation}
 \label{Eq:VdropSimple}
V_{drop} = \Delta V - \frac{\Delta V}{1 + R_{bias} \cdot F_{pe} \cdot C_{\mu cell}}
\end{equation}
where $F_{pe}$ is the CL rate expressed in p.e. per second.

To have a precise calculation of $V_{drop}$ and at the same time avoid the complexity of solving Eq.~\ref{Eq:SiPMCurrent} analytically, a toy MC model is developed.

\section{The toy Monte Carlo \label{sec:toy}}

The described model in Sec.~\ref{sec:model} is implemented into a toy MC.
In the first step, all relevant SiPM parameters are measured experimentally for the large area ($\sim$1~cm$^2$) hexagonal SiPM, S10943-2832(X), produced by Hamamatsu HPK \cite{Hamamatsu} for the single mirror small size telescope SST-1M camera. All these parameters and their measurements are described in Ref.~\citep{Nagai:2018ovm}.

Each simulated time interval, typically between 200 and 2'000 ns, was sampled with a given sampling rate $R_{s}$ and sample time width $\Delta t_{i} = 1/R_{s}$ (typically, in the interval 100~ps~$\leq$~$\Delta t_{i}$~$\leq$~4~ns). For each sample $i$, in the range from 0 to $S-1$, the main simulation steps are:
\begin{enumerate}
    \item randomly generate a number of photons $N_{gen}(\lambda, i)$ using a Poisson distribution with mean value of $F_{ph}(i) \cdot \Delta t_{i}$ according to the CL rate with a given wavelength distribution or single wavelength.
   If the simulated optical system contains wavelength filters, e.g. entrance window of the camera with anti-reflective coating and low pass filer, it can be accounted for at this stage;
    \item each generated photon is processed separately. It may be detected or not, depending on the $PDE$ and photon wavelength:
    \begin{equation}
        N_{light}^{p.e.}(i) = \sum_{n=1}^{N_{gen}(\lambda, i)} f(r_{n})
    \end{equation}{}
    where $r$ is a random number uniformly distributed from 0 to 1 and $f(r)$ is:
    \begin{equation}
        f(r) = \begin{cases}
    1       & \quad \text{if } r \leq PDE(\Delta V(i), \lambda)\\
    0  & \quad \text{if } r > PDE(\Delta V(i), \lambda)
  \end{cases}
    \end{equation}{}
    \item on top of the CL (i.e. $N_{gen}(\lambda, i)$), the SiPM uncorrelated noise (with rate $DCR$) is added:
    \begin{equation}
        N_{tot}^{p.e.}(i) =  N_{light}^{p.e.}(i) + N_{DCR}(i, r)
    \end{equation}
    where $N_{DCR}$ is the number of dark pulses, calculated as:
    \begin{equation}
        N_{DCR}(i, r) = \begin{cases}
    1       & \quad \text{if } r \leq DCR(i, \Delta V) \cdot \Delta t_{i} \\
    0  & \quad \text{if } r > DCR(i, \Delta V) \cdot \Delta t_{i}
  \end{cases}
    \end{equation}{}
    Here, we neglect that two or more dark pulses may appear within the same $\Delta t_{i}$, because even for large $\Delta t_{i}$ of 4 ns this probability is less than 1\%. 
    
    \item $N_{tot}^{p.e.}(i)$ are randomly enhanced by optical crosstalk and/or afterpulses. The total number of avalanches for a given sample $i$, is then: $N_{av}(i)$;
    
    \item randomly create an avalanche generation time $t_{av}$ from an uniform distribution in the range 10 ps + ($i\times \Delta t_{i}$) $\leq$ $t_{av}$ $<$ $\Delta t_{i} \times (i + 1)$; 
    
    \item $N_{av}(i)$ is converted into the SiPM current $I_{SiPM}(i)$ as:
    \begin{equation}
        I_{SiPM}(i) = F_{av}(i) \times \frac{C_{\mu cell}
        \cdot \Delta V (i) }{e} = \frac{N_{av}(i)}{\Delta t_{i}} \times G(\Delta V(i)).
    \end{equation}{}
    In this step $G(\Delta V(i))$ is randomly smeared with a Gaussian distribution with standard deviation $\sigma_{G}$ corresponding to the SiPM gain fluctuation between different micro-cells;
    \item $I_{SiPM}(i)$ is used to calculate $V_{drop}(i)$, and both are used to estimate the effect on  the overvoltage in the sample $i+1$:
    \begin{equation}
        \Delta V (i+1) = \Delta V (i) - V_{drop} (i)
    \end{equation}
    The overvoltage $\Delta V (i+1)$ is used to derive the values of all parameters (e.g. $G$, $A_{p.e.}$, $PDE$, $DCR$, $P_{ct}$ and $P_{ap}$) for the sample $i+1$;
    \item the arrival time of detected and generated photons, as well as all parameters used in the simulation, are stored in a ROOT\footnote{\url{https://root.cern.ch}} binary file for future use. 
\end{enumerate}
At the last sample $i=S-1$, the experimental waveform is generated as the sum of all the generated avalanches $N_{av}(i)$ convoluted with the template of the typical normalized SiPM pulse with its amplitude $A_{p.e.}(\Delta V(i), i)$ and shifted by the initial electronic baseline. Additionally, each waveform value is randomly smeared by a Gaussian distribution with a standard deviation $\sigma_{e}$ corresponding to the electronic noise of the system under consideration.

\begin{figure}[hbt]
    \begin{center}
        \includegraphics[width=0.8\textwidth, trim=0 2.cm 0 0, clip]{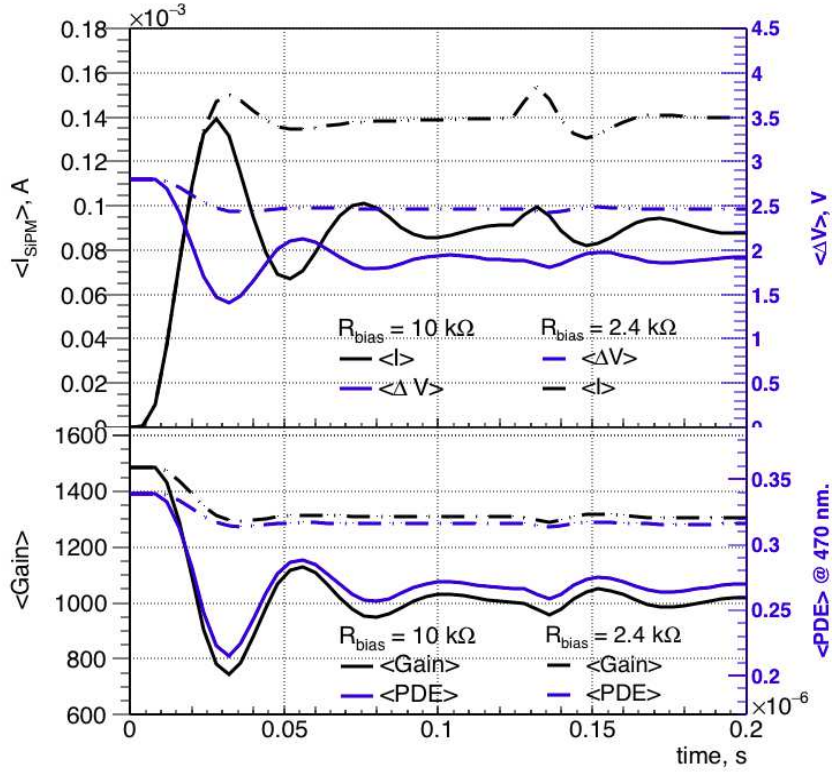}
    \end{center}
    \caption{The simulated current in a SiPM, $I_{SiPM}$, vs time under CL with photon rate of 2~GHz. $I_{SiPM}$ induces a drop of $\Delta V$, which in turn affects all main SiPM parameters (i.e. $G$, $PDE$, $P_{ct}$ , $P_{ap}$ and $DCR$). As an example the $PDE$ and $G$ as a function of time are presented in the bottom plot. Results are presented for two values of $R_{bias}$ of 10~k$\Omega$ and 2.4~k$\Omega$.}
    \label{Fig:ParamVsTime}
\end{figure}

To illustrate the simulation steps, the average $I_{SiPM}$ resulting from the simulation of a SiPM illuminated with a photon rate of $F_{ph}$~=~2~GHz as a function of time is presented in Fig.~\ref{Fig:ParamVsTime} (top) for two values of $R_{bias}$ of 2.4~k$\Omega$ and 10~k$\Omega$. Relatively fast changes of the main SiPM parameters, in particular of the over-voltage $\Delta V$, and consequently of the $PDE$, are observed within the first time steps of the simulation before the steady state is reached. Such a behaviour is related to recursive conjugation of $I_{SiPM}$ and $\Delta V$. At $t=0$, $I_{SiPM}$ increases with $N_{av}$ (see Eq.~\ref{Eq:SiPMCurrent}), but it is quickly quenched by the presence of the bias resistor which causes the over-voltage to decrease (See Eq.~\ref{Eq:OverVoltageWithDrop}). The time interval before the steady state is achieved increases with $F_{ph}$ and $G$. For this particular example, it is reached after $\sim$100~ns.


\section{Validation of the toy Monte Carlo \label{sec:valid}}

The proposed toy MC is compared with a simplified analytical calculation~\citep{SST1Melectronics}, and then with measurements obtained in the laboratory with a calibrated light source.
It is then compared with data taken with the SST-1M camera and its Camera Test Setup (CTS) \cite{CameraPaperHeller2017}. The results are described below.

\subsection{Validation with the analytical calculation}

\begin{figure}[hbt]
    \begin{center}
        \includegraphics[width=0.7\textwidth
        ]{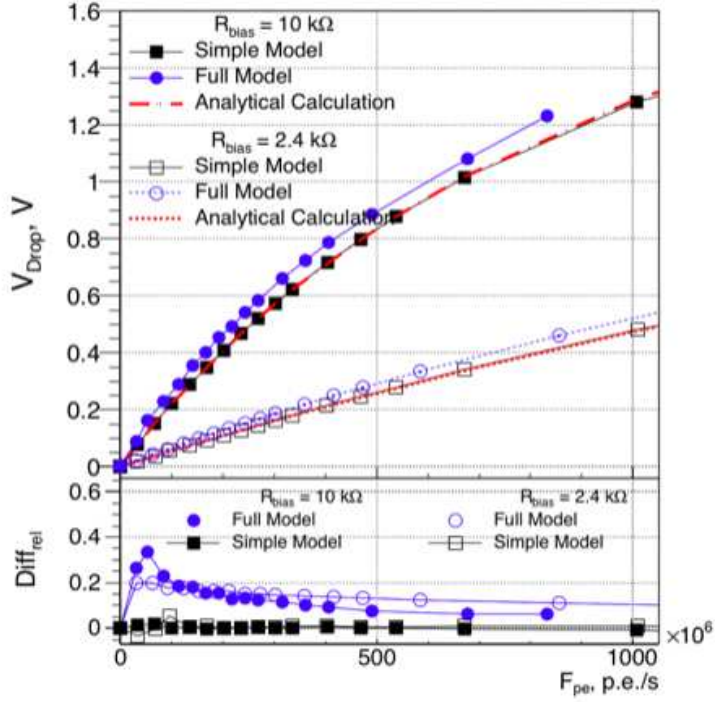}
    \end{center}
    \caption{$V_{drop}$ as a function of $F_{pe}$, calculated analytically (red line), with the simplified toy MC (which assumes $PDE$ = 100$\%$, $DCR$ = 0, $P_{ap}$ = 0, $P_{ct}$ = 0) (black squares) and with the toy MC with real SiPM parameters ($PDE(\Delta V)$, $P_{ct}(\Delta V)$, $P_{ap}(\Delta V)$ and $DCR(\Delta V)$) (blue circles). Results are for two values of $R_{bias}$ of 10~${\rm k}\Omega$ (closed symbols) and 2.4~${\rm k}\Omega$ (open symbols). 
    }
    \label{Fig:VdropVsNSBSimpleModel}
\end{figure}

As discussed in Sec.~\ref{sec:model}, the analytical calculation of voltage drop can be done assuming that CL affects only the SiPM gain $G$. To compare this analytic expression with the proposed toy MC, the voltage drop is simulated using real SiPM parameters and using a simplified case. For this simplified case, we use the toy model assuming null values for $DCR$, $P_{ct}$ and $P_{ap}$ and that the $PDE$ is 100\%. The results for $R_{bias} = 10~{\rm k}\Omega$ (full symbols) and 2.4~${\rm k}\Omega$ (empty symbols) are presented in Fig.~\ref{Fig:VdropVsNSBSimpleModel}. Independently of $R_{bias}$, the comparison between the analytical expression (lines) and the simplified model (squares) shows an excellent agreement. The relative difference is less than 0.5\% on average. When compared to the full model (circles), the relative difference increases to an average of $\sim$14\%, which is expected due to the assumed simplifications to use the analytical expression.


\subsection{Validation with calibrated light sources}
\label{Sec:CalibratedSource}
\begin{figure}[hbt]
    \begin{center}
        \includegraphics[width=1\textwidth, keepaspectratio]{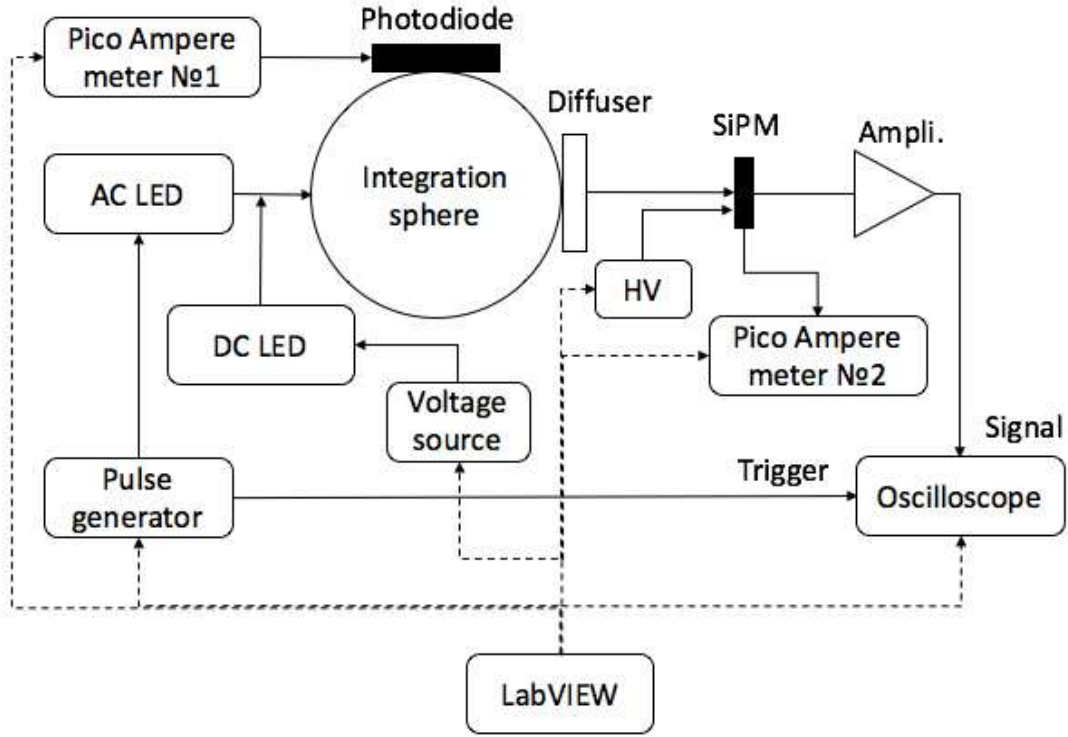}
    \end{center}
    \caption{Schematic layout of the experimental set-up used for the toy model validation. Two LEDs are used, one in AC mode to emulate the signal and another in DC mode to emulate various CL levels. The light intensity is measured by the calibrated photodiode. An additional pico-amperometer is used to directly measure the voltage drop and an oscilloscope is used for data acquisition.}
    \label{Fig:Schematic}
\end{figure}

\begin{figure*}[hbt]
\centering
\includegraphics[width=0.5\textwidth]{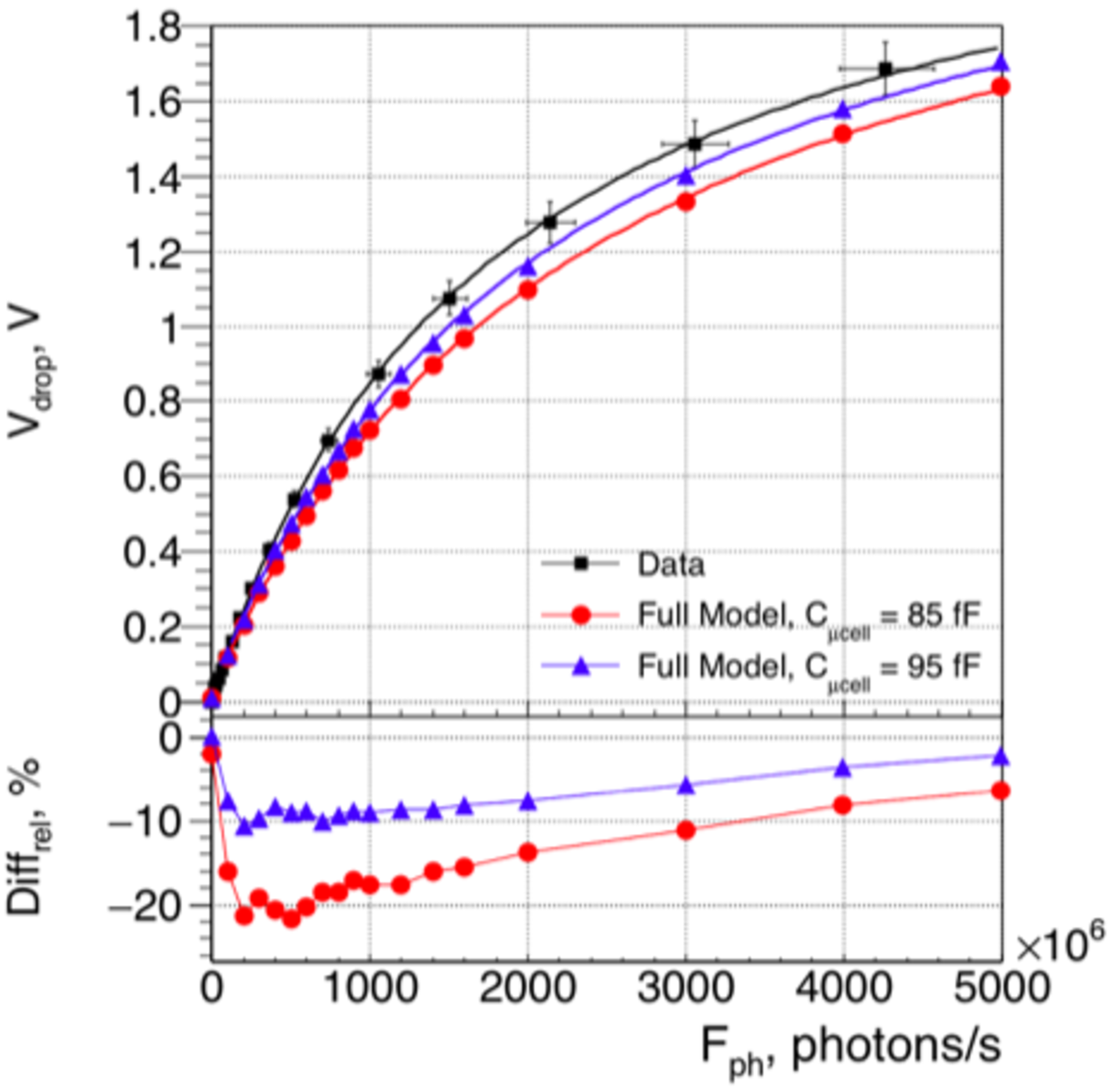}\hfill
\includegraphics[width=0.5\textwidth]{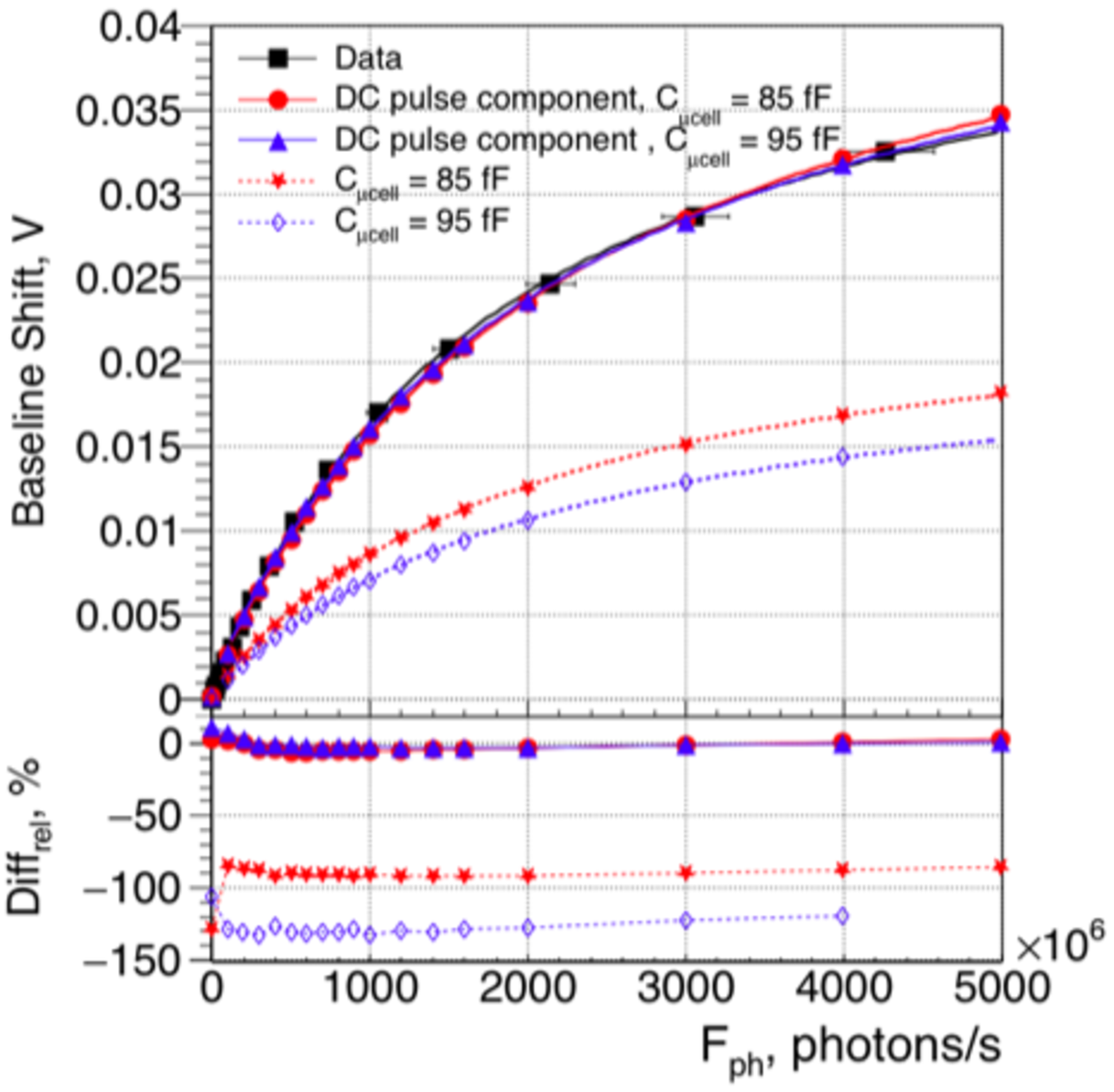}
\caption{Voltage drop (left) and baseline shift (right) as a function of the CL level in photons per second. Data (black rectangles) are compared to the toy MC assuming two values of $C_{\mu cell}$ of 85~fF (red) and 95~fF (blue). This 85~fF capacitance value was provided by the producer, but we also use 95~fF because of possible 10\% parasitic capacitance. The results for the baseline shift from the toy MC is presented with (solid lines) and without correction (dashed lines) for the DC pulse component.}
\label{Fig:VdropAndBaselineVsNSB}
\end{figure*}

For this study, the experimental setup at IdeaSquare\footnote{\url{http://ideasquare.web.cern.ch}} at CERN was used, and is illustrated in Fig.~\ref{Fig:Schematic}. The full description of the set-up may be found in \citep{Nagai:2018ovm}. The SiPM  is biased with a Keithley 2410 through an RC filter ($R_{bias}$ = 10~k$\Omega$, $C_{bias}$ = 100~nF). The SiPM anode is connected to the Keithley Picoammeter 6487 to measure the bias voltage at the SiPM and also to the preamplifier board developed for the SST-1M camera (more details can be found in \citep{SST1Melectronics,CameraPaperHeller2017}). The waveform readout is performed with a Lecroy 620Zi oscilloscope. The set-up is equipped with two LEDs ($\lambda$ = 470~nm each). A LED is pulsed in $AC$ mode (to emulate the flashes of Cherenkov light induced by atmospheric showers), while the other is continuous, or in the so-called $DC$ mode (to emulate the NSB or CL). The SiPM under study was biased to  $V_{PS}$ = 58~V and temperature of 25~$^{\circ}{\rm C}$, corresponding to $\Delta V = 3.22$~V. For each light level, 10'000 waveforms are acquired, each of 2~$\mu$s long (5'000 samples per waveform). The light intensity is monitored with $5\%$ precision using a calibrated photodiode\footnote{Hamamatsu S1337-1010BQ, s/n 61}. 
Two types of measurements were performed, as described next. 
\paragraph{\textbf{DC scan}} The first measurement is performed with different CL levels. The LED in continuous mode is used to emulate the various CL levels. The measured $V_{drop} = V_{PS} - V_{SiPM}$ and average waveform baseline are compared with simulated values. Fig.~\ref{Fig:VdropAndBaselineVsNSB}-left shows a good agreement between the measured and the simulated $V_{drop}$ as a function of $F_{ph}$ in terms of photons/s (the number of photons can be extracted using the calibrated pohotodiode). However, the measured baseline shift is almost two times higher with respect to the simulated one (see Fig.~\ref{Fig:VdropAndBaselineVsNSB}-right).

The difference in baseline shift can be explained by the slow pulse component (referred to as ``DC pulse component''), which extends over 150 ns. This DC pulse component is not included directly in the toy model, as the adopted pulse template is only 150~ns long. The DC pulse component is difficult to measure because of the high $DCR$ and $P_{ap}$ in this large SiPM area (93.6~mm$^{2}$). Therefore, the DC pulse component is measured using low rates of injected light ($F_{ph} < 120$~MHz). This light level is chosen to fulfill two requirements:
\begin{itemize}
    \item single SiPM pulses are still distinguishable~\footnote{A single pulse is a SiPM signal separated by the neighboring pulses by a time interval higher than $\Delta t_{length} = \Delta t_{rec.} + \Delta t_{rise} + \Delta t_{baseline}$, where $\Delta t_{rec.}$ and $\Delta t_{rise}$ are the typical SiPM recovery and rise times, respectively, and $\Delta t_{baseline}$ = 40~ns is the time interval during which the local baseline calculation is performed.};
    \item the $PDE$, used to convert $F_{ph}$ into $F_{pe}$ can be considered constant\footnote{We estimated a relative $PDE$ drop of 1.75 $\%$ for $F_{ph}$ = 120 MHz.}.
\end{itemize}{}
For each pulse, the local baseline is calculated within 40~ns before the pulse. The deviation of the local baseline, at a given $F_{ph}$ level, from its value in the dark is called in the following ``local baseline shift'': $V_{BLS}$. The $V_{BLS}$ is converted from a voltage into a charge as:
\begin{equation}
    Q_{BLS} = \frac{V_{BLS} \cdot \Delta t}{R_{load}} 
\end{equation}
where $\Delta t$ is the waveform duration and $R_{load} = 50~\Omega$ is the load resistance. At the same time, $F_{ph}$ is converted from number of photons into SiPM detected charge as:
\begin{equation}
    Q_{ph} = F_{ph}(\lambda) \cdot PDE(\Delta V, \lambda) \cdot \Delta t \cdot \frac{C_{\mu cell} \cdot \Delta V}{e}
\end{equation}

\begin{figure}[hbt]
    \begin{center}
        \includegraphics[width=0.6\textwidth]{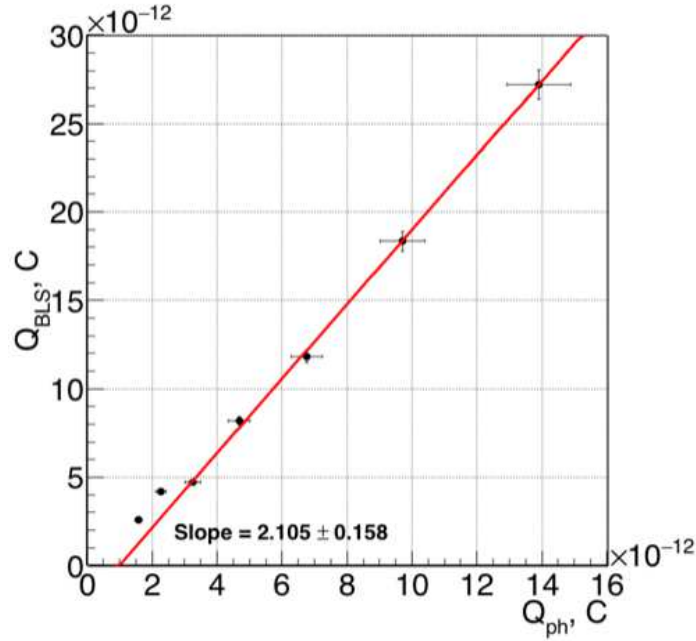}
    \end{center}
    \caption{DC pulse component, expressed in baseline shift charge $Q_{BLS}$ as a function of generated SiPM charge $Q_{ph}$, due to CL.}
    \label{Fig:LocalBaselineVsNSB}
\end{figure}

\begin{figure}[hbt]
    \centering
    \includegraphics[width=0.45\textwidth]{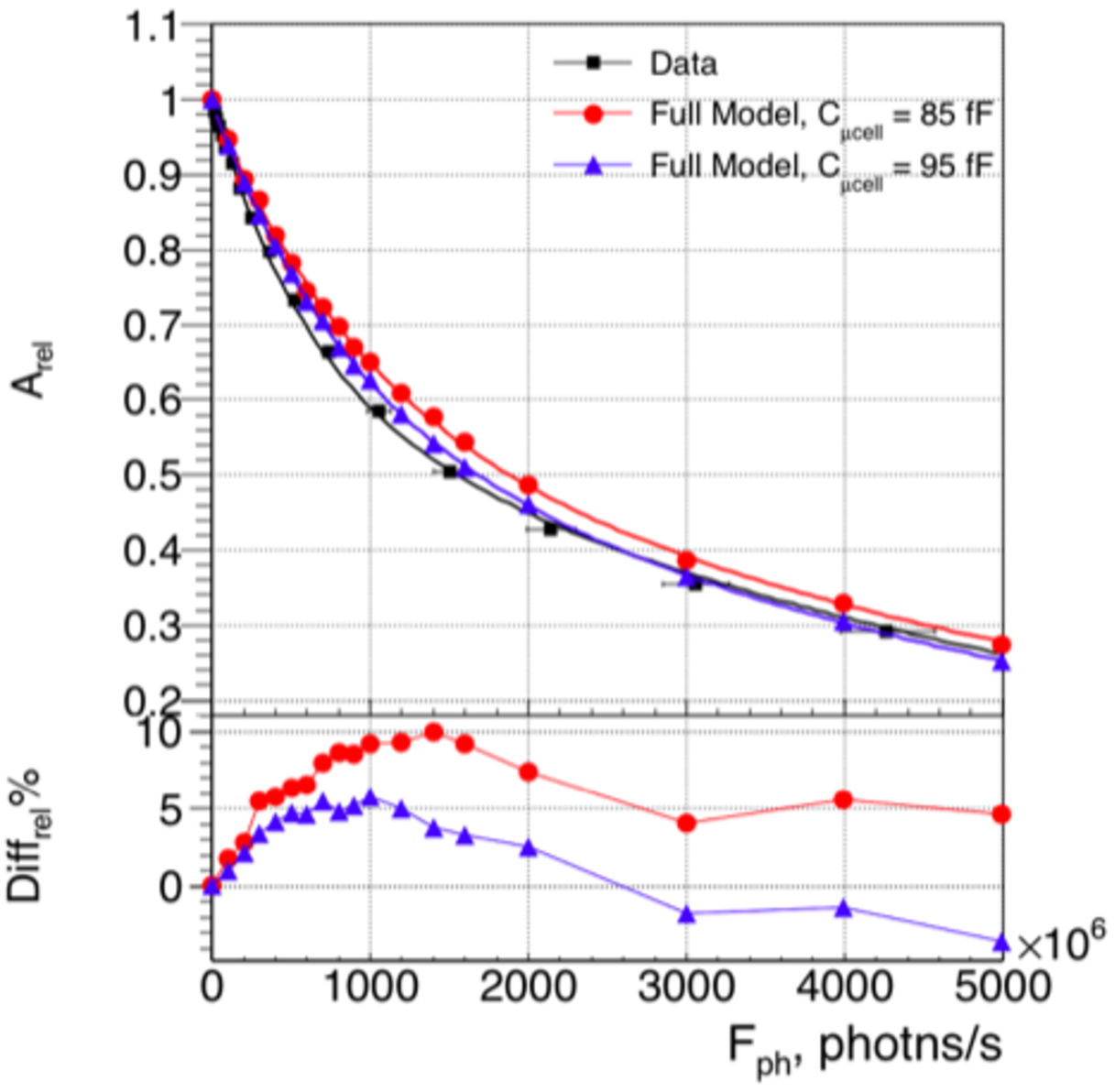}
    \includegraphics[width=0.45\textwidth]{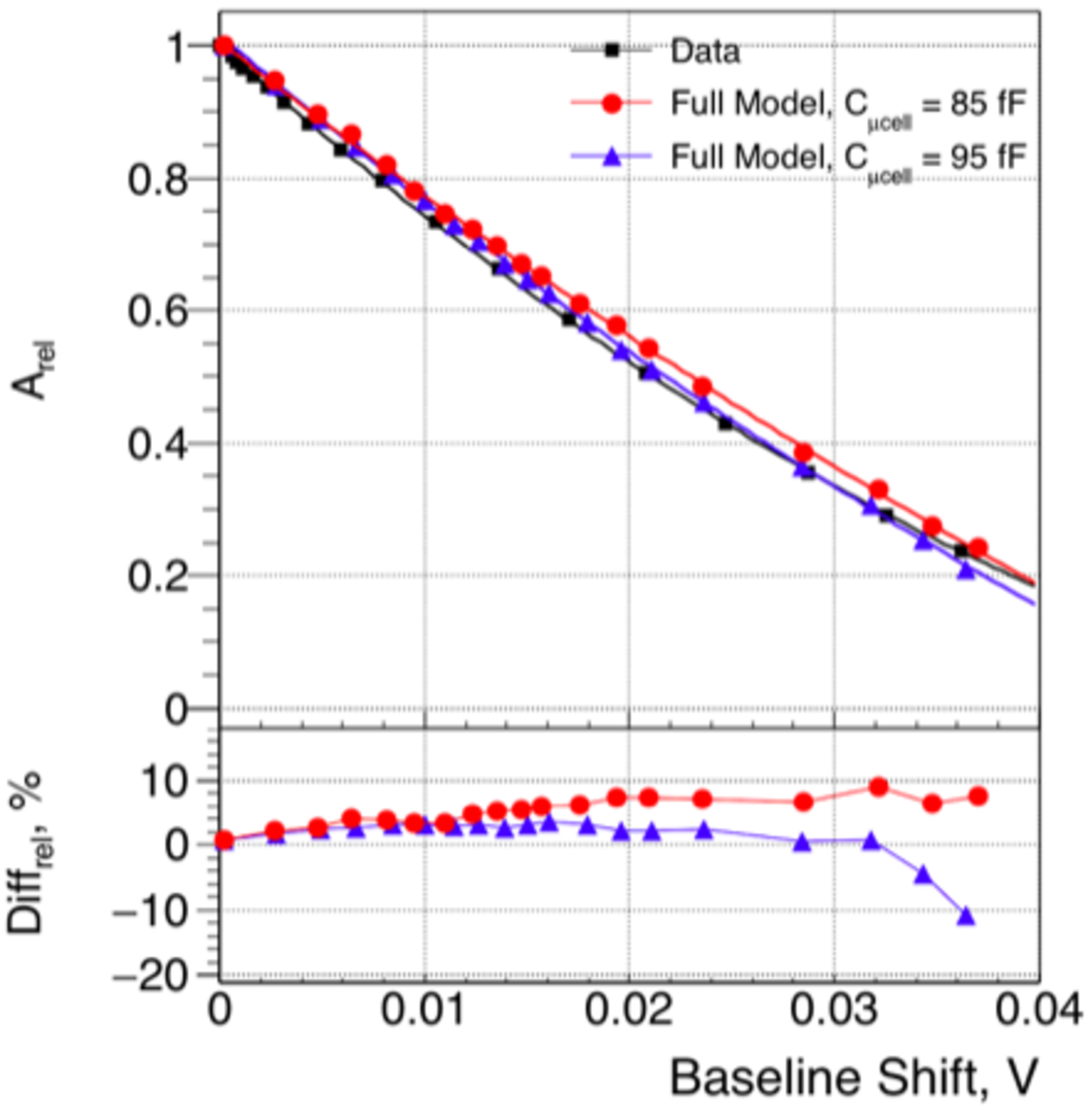}
    \caption{Relative amplitude $A_{rel.}^{Signal}$ from measurements in the laboratory with calibrated light sources (black squares) and from proposed toy MC simulation (red circles for  $C_{\mu cell} = 85$~fF and blue triangles for 95 fF) as a function of CL photon rate (left) and baseline shift (right). The relative difference in the relative amplitude for data and simulation is presented in the bottom plots. }
\label{Fig:ArelVsNSBLab}
\end{figure}

The average $Q_{BLS}$ as a function of the $Q_{ph}$ is presented in Fig.~\ref{Fig:LocalBaselineVsNSB}. We can see that the $Q_{BLS}$ increases linearly with $Q_{ph}$ with a slope of $BL_{slope} = \ 2.1$. This slope indicates that only 32.3\% of the charge generated by the SiPM is seen as a pulse, while the remaining 67.7\% goes into the baseline shift. Implementing this behaviour inside the toy MC dramatically increases the simulation time. Therefore, the effect is accounted for some additional steps after the simulation is performed. As a matter of fact, each simulated waveform is shifted by an additional baseline $BL_{add}$, calculated as:

\begin{equation}
    BL_{add} = R_{load} \cdot BL_{slope} \cdot \frac{\sum I_{SiPM}}{\Delta t},
\end{equation}

where $\sum I_{SiPM}$ is the total current generated by the SiPM within the simulated waveform. The results obtained accounting for this additional baseline shift are presented in Fig.~\ref{Fig:VdropAndBaselineVsNSB} (right) with solid lines and indicated as ``DC pulse component'' in the legend.
\paragraph{\textbf{AC/DC scan}} The second measurement is performed with different pulsed and CL levels. 
The average waveform amplitude, $A$, and average waveform baseline, $BL$, corresponding to the AC LED, depend on $F_{ph}$ due to the voltage drop effect. The variation with respect to the case without CL can be calculated as: 
\begin{equation}
    A_{rel} = \frac{A(F) - BL(F) }{ A(F0) - BL(F0) }
\label{Eq:RelativeAmpli}
\end{equation}
where $A(F)$ and $BL(F)$ are the average amplitude and average baseline, respectively, at a given DC LED intensity, while $A(F0)$ and $BL(F0)$ are corresponding values at zero DC intensity.
As shown in Fig.~\ref{Fig:ArelVsNSBLab}-left, we can observe that the detected pulse amplitude after baseline subtraction decreases with increasing DC LED intensity, as it is expected due to $V_{drop}$. This behaviour was compared with the results from the toy model for $C_{\mu cell}$ of 85 and 95~fF. We can observe that the maximum difference between measured and simulated values, for $C_{\mu cell} = 95$~fF, is less than 5\%.

Typically, during the operation of SiPMs in real conditions, the CL level is unknown. However, it can be calculated from the baseline shift or its standard deviation. For these purposes the relative amplitude, $A_{rel.}$, as a function of the baseline shift is presented in Fig. \ref{Fig:ArelVsNSBLab}-right.

\subsection{Validation with the Camera Test Setup}
\label{Sec:ValidCTS}

\begin{figure}[hbt]
    \centering
    \includegraphics[width=0.6\textwidth]{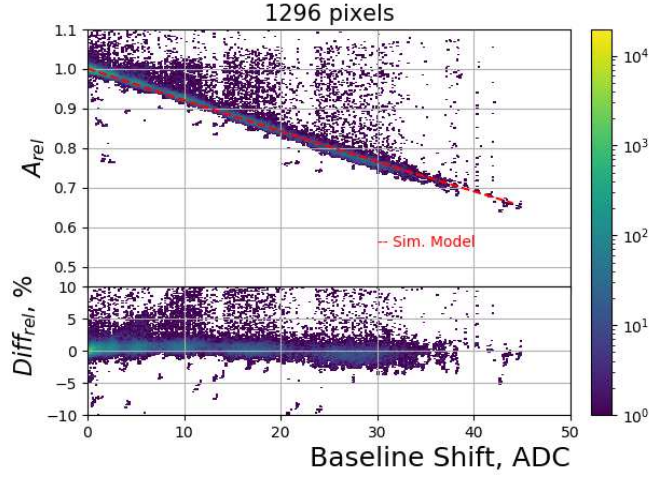}
    \caption{DC/AC scan using 12 different continuous light levels ($DC$ LED DAC) and injecting for each CL 19 $AC$ LED DAC level pulses. The relative amplitude  $A_{rel}$ for all 1296 camera pixels is shown as a function of Baseline shift. 
    The $A_{rel}$ from the proposed model in this paper is represented by the dashed red line. The relative difference in the relative amplitude for data and simulation is presented in the bottom plots.}
    \label{Fig:AmpliDropCTS}
\end{figure}

Further measurements are performed with the SST-1M camera~\citep{CameraPaperHeller2017} and its camera test setup (CTS),
 at the University of Geneva. The CTS calibration tool, described in Ref.~\cite{CameraPaperHeller2017}, is equipped with two LEDs ($\lambda$ = 468~nm) corresponding to each SST-1M camera pixel: one in pulsed mode ($AC$ LED) and the other in continuous mode ($DC$ LED). With the CTS, the $AC/DC$ scan described in Sec.~\ref{Sec:CalibratedSource} is done for all 1296 camera pixels at $\Delta V$ = 2.8 V. For each pixel, AC and DC LED values, the baseline and signal amplitude are calculated. The relative amplitude (see Eq.~\ref{Eq:RelativeAmpli}) as a function of the baseline shift
is presented in Fig.~\ref{Fig:AmpliDropCTS}.
We can observe that almost for all pixels $A_{rel.}$ decreases with increasing baseline shift (i.e. CL).
Few pixels do not follow this tendency as either the pixel itself or the LEDs facing it were found to be faulty. In addition, some LEDs have a higher intensity with respect to others resulting in the saturation of the pixel readout chain~\citep{SST1Melectronics}. Therefore, no drop of  $A_{rel.}$ is observed. The relative difference between data and the proposed model is shown in Fig.~\ref{Fig:AmpliDropCTS} (bottom). It is around 5\%, 
confirming the results shown in the previous section.

\section{Results \label{sec:results}}

\begin{figure*}[hbt]
    \centering
    \includegraphics[width=0.32\textwidth]{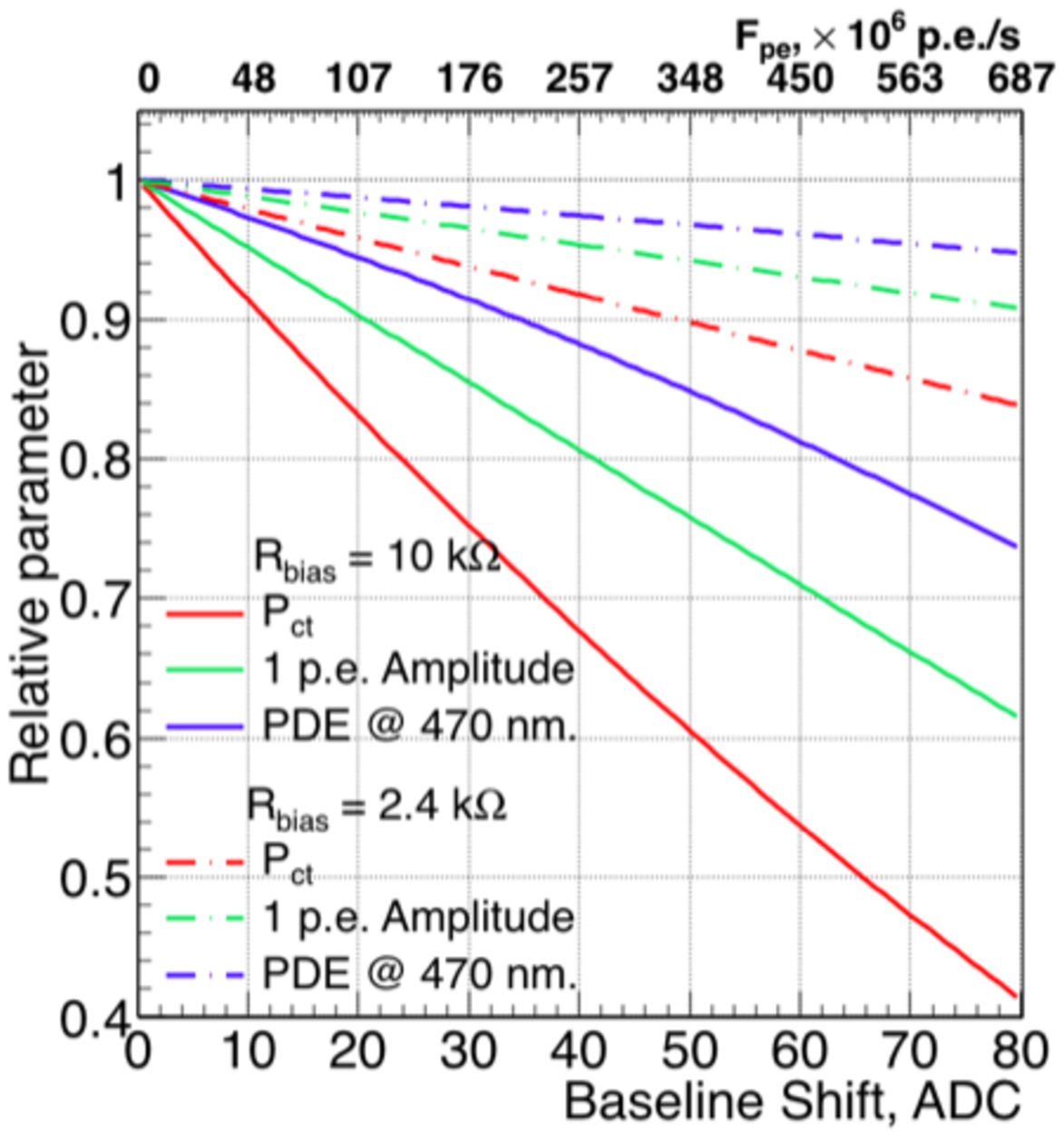}
    \includegraphics[width=0.33\textwidth]{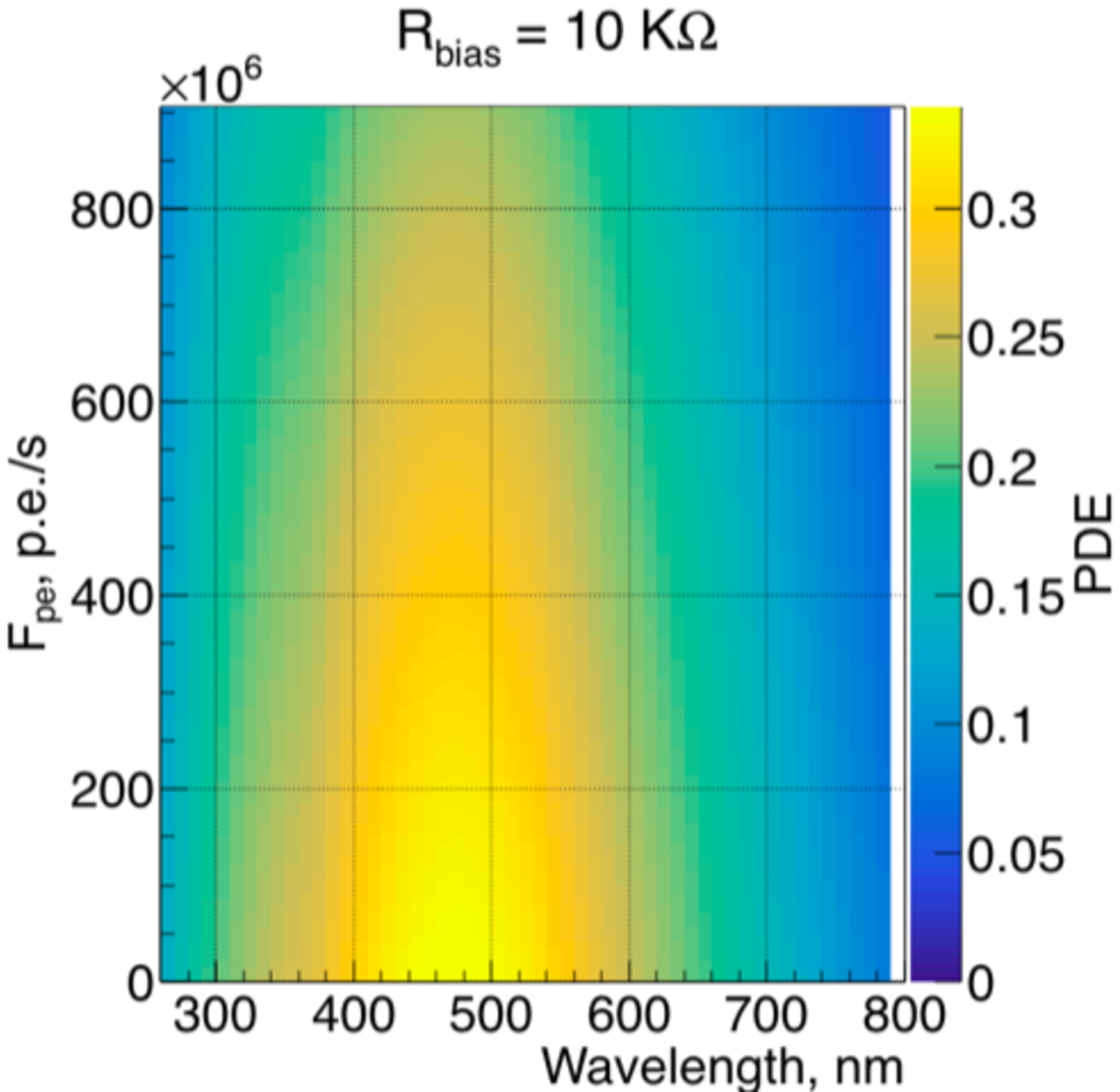}
    \includegraphics[width=0.33\textwidth]{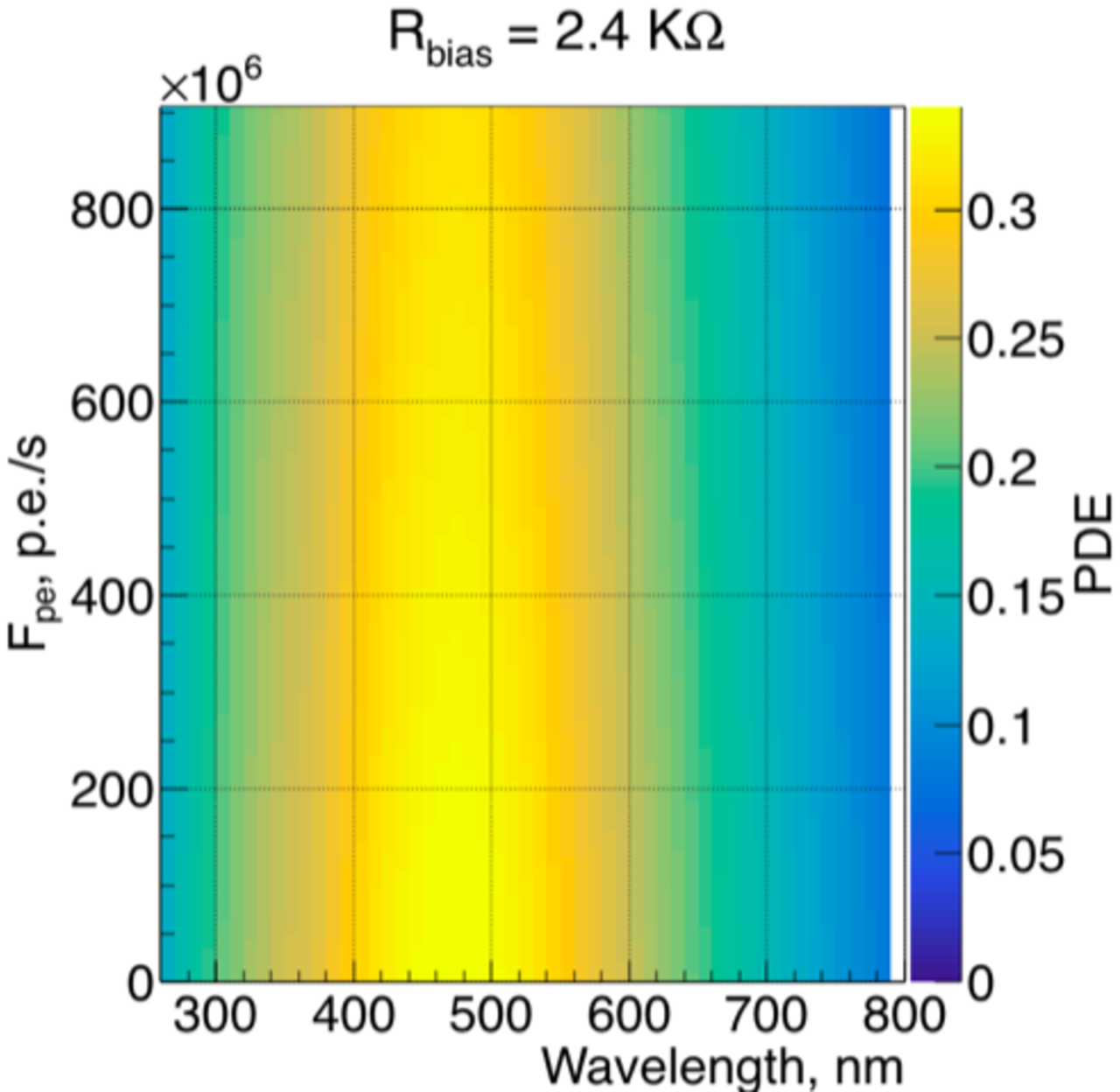}
    \caption{Top: Relative 1 p.e. Amplitude, $P_{ct}$ and $PDE$ as a function of CL rate (in p.e./s) and wavelength for two values of $R_{bias}$ of 10 k$\Omega$ and 2.4 k$\Omega$. The middle and right plots show for the two indicated $R_{bias}$ values the PDE as a function of the CL rate in p.e. and wavelength.}
    \label{Fig:PDEDropBothRbias}
\end{figure*}

From the proposed toy model, the CL level and the $V_{drop}$ can be obtained from the baseline shift or its standard deviation.
Therefore, all SiPM parameters can be corrected according to $V_{drop}$, as shown in Fig.~\ref{Fig:PDEDropBothRbias} for the $PDE$, $P_{ct}$ and the amplitude of the single p.e. signal. By comparing the relative drop of the main SiPM parameters with CL for $R_{bias}$ of 10 k$\Omega$ and 2.4 k$\Omega$ we can conclude that for $R_{bias}$ of 2.4 k$\Omega$ the relative drop is more than three times smaller, because $V_{drop}$ is proportional to $R_{bias}$ (See Eq.~\ref{Eq:OverVoltageWithDrop}). Hence, low values of $R_{bias}$ simplify operation of SiPM under CL.
A similar plot can be obtained in photons/s when the effect of $PDE$ and additional optical elements (light guides, window, etc.) are included in the simulation.


\begin{figure}[hbt]
    \begin{center}
       \includegraphics[width=0.49\textwidth]{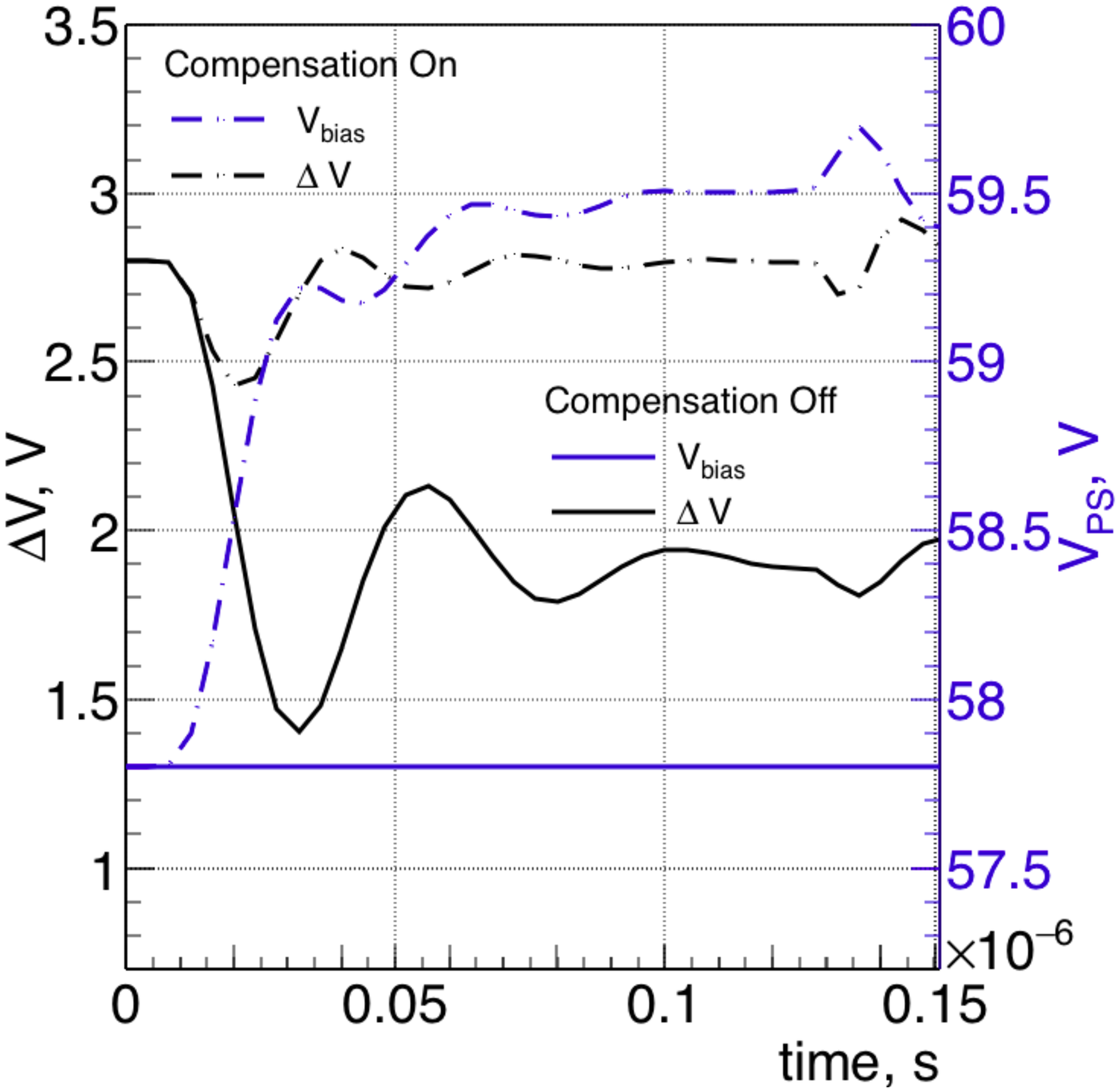}
       \includegraphics[width=0.49\textwidth]{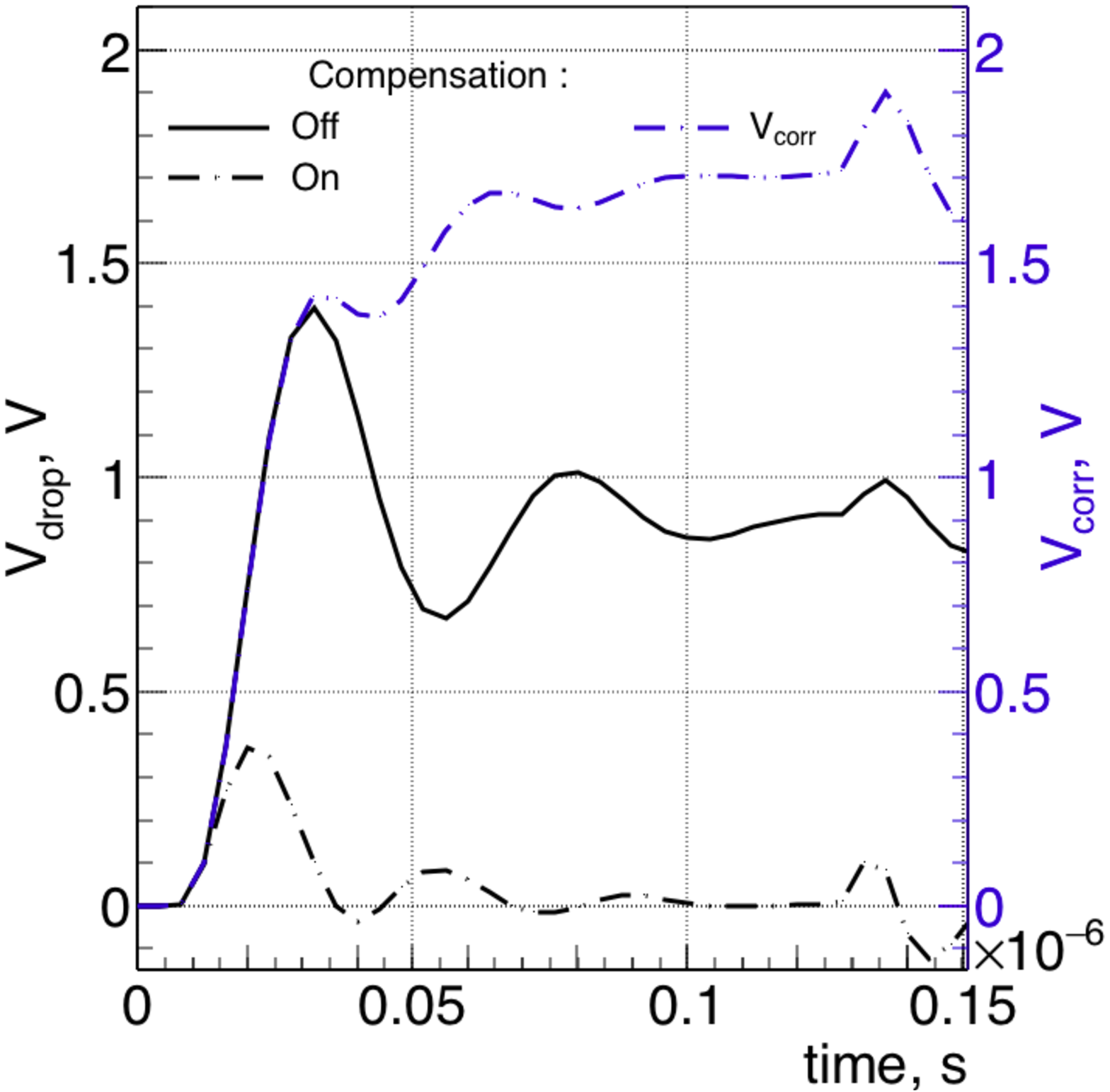}
    \end{center}
    \caption{Evolution of over-voltage $\Delta V$ and $V_{PS}$ averaged over 10'000 samples with time under a CL of $2\times 10^{9}$ photons/s, with and without compensation loop (left). The voltage drop $V_{drop}$ and the bias voltage correction $V_{corr}$ are presented at the right.}
\label{Fig:VoltageCorrection}
\end{figure}

The drop of the SiPM parameters (See Fig.~\ref{Fig:PDEDropBothRbias}) under CL may be compensated by increasing the bias voltage $V_{SiPM}$ by some correction voltage $V_{corr}$ in order to keep constant the over-voltage $\Delta V$ (see Eq.~\ref{Eq:OverVoltageWithDrop}). We call this ``compensation loop''. 
$V_{corr}$ is determined from the toy MC. As an example, the evolution of $\Delta V$ and $V_{drop}$ with time under CL of $2\times 10^{9}$ photons/s is shown with compensation loop (dashed lines) and without (solid lines) in Fig.~\ref{Fig:VoltageCorrection} (top). We can observe that, to compensate by $V_{drop} \sim 0.9$~V, the $V_{PS}$ should be increased by 1.7~V, as shown in Fig.~\ref{Fig:VoltageCorrection} (bottom). As a drawback, the detected NSB rate increases from $505\times 10^{6} \ p.e./s$ up to $653\times 10^{6} \ p.e./s$. Hence, the SiPM power consumption increased from 5.17~mW up to 10.29~mW. 

\begin{figure}[hbt]
    \begin{center}
        \includegraphics[width=0.49\textwidth]{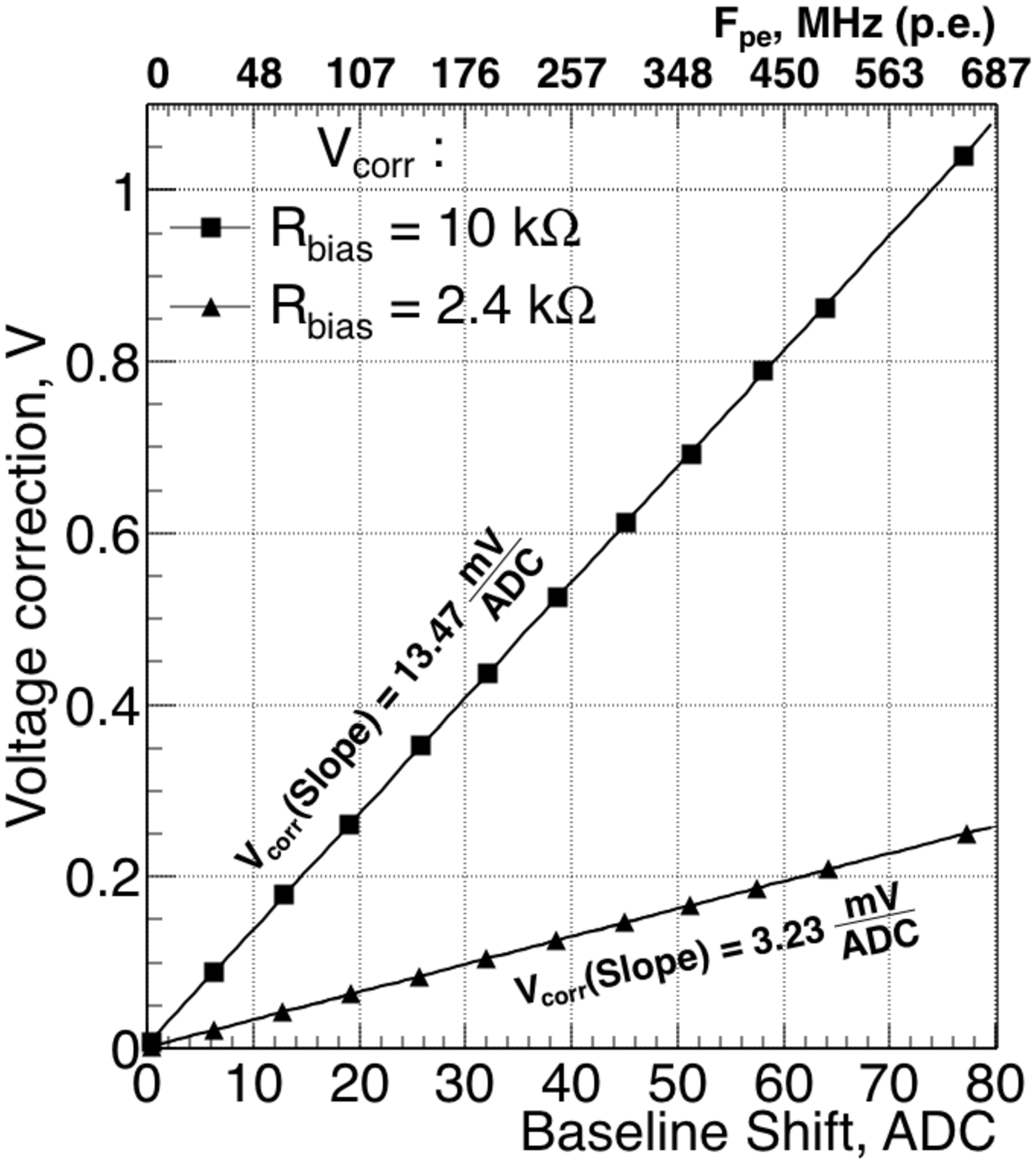}
        \includegraphics[width=0.49\textwidth]{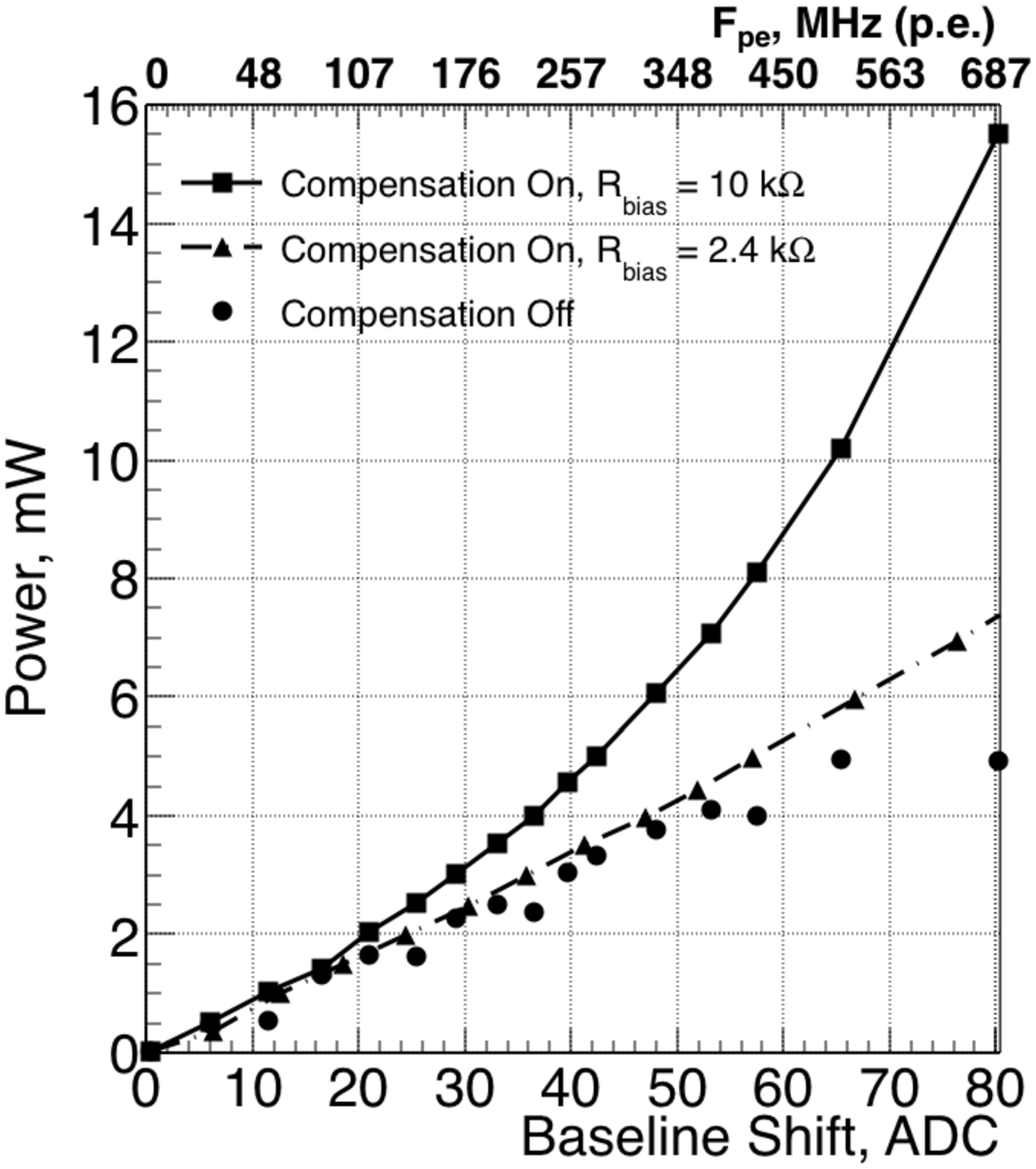}
    \end{center}
    \caption{Correction bias voltage $V_{corr}$ (left) that needs to be applied to have a constant over-voltage of $\Delta V$=2.8~V as a function of the baseline shift for two values of $R_{bias}$ of 10 $k\Omega$ (rectangles) and 2.4 $k\Omega$ (triangles). A linear behaviour of $V_{corr}$ with the baseline shift is found with slope of 13.47 ($R_{bias} = 10 ~k\Omega$) and 3.23~mV/$ADC$ ($R_{bias} = 2.4 ~k \Omega$). The sensor power consumption is also shown (right)
    On the right y-axis (in bluw) the sensor power consumption is shown with the same symbols for the two bias resistors as before. The circles are the SiPM power consumption with no compensation loop.}
\label{Fig:VbiasCorrection}
\end{figure}

The $V_{corr}$ as a function of baseline shift is presented in Fig.~\ref{Fig:VbiasCorrection} for two values of $R_{bias}$ of 10~k$\Omega$ and 2.4~k$\Omega$. We can observe that $V_{corr}$ increases linearly with increasing baseline shift with a slope of 13.47~$\frac{mV}{ADC}$ and 3.23~$\frac{mV}{ADC}$ for  $R_{bias}$ of~10 and 2.4~k$\Omega$, respectively. Therefore, in experimental conditions, when CL is known and stable in time, the effects from $V_{drop}$ can be corrected. As a drawback, the SiPM power consumption increases after compensation loop activation, as shown in Fig.~\ref{Fig:VbiasCorrection}. Also, the baseline standard deviation $BL_{\sigma}$ can be used to calculate $V_{corr}$. However, it is less precise since $BL_{\sigma}$ shows stronger dependence on statistics and  electronic noise.

\section{Conclusions}

In this paper we report on the studies of SiPM behaviour under CL. A Toy Monte Carlo model was developed for DC coupled electronics. This model is used to predict the behaviour of all relevant SiPM parameters (i.e. Gain, Photon detection efficiency, optical crosstalk, after-pulses, dark count rate and etc.) under various CL. The model is validated by comparison with experimental data measured for a single SiPM as well as for the full SST-1M gamma-ray camera, which contains 1296 SiPM devices.
This model can be adapted to any DC coupled SiPM. Indeed, it can also be extended to the AC coupling case. As a matter of fact, 
in~\cite{CameraPaperHeller2017} it can be seen that the standard deviation of the waveform also increases with increasing CL. This parameter can therefore be used as an indicator for AC coupled systems, similarly to the baseline shift that we used for DC coupling.
However, we showed that a DC couple system is preferable as the standard deviation tends to saturate at large CL levels while the baseline shift does not.


\bibliography{BibFile}

\end{document}